\documentclass[11pt,conference]{IEEEtran}
\usepackage{cite}
\usepackage{amsmath,amssymb}
\usepackage{graphicx}
\usepackage[hidelinks]{hyperref}

\title{\fontsize{18}{20}\selectfont Agentic Semantic Control for Autonomous Wireless Space Networks: Extending Space-O-RAN with MCP-Driven Distributed Intelligence}

\author{
    Eduardo Baena, Paolo Testolina, Michele Polese, Sergi Aliaga, Andrew Benincasa,\\ 
    Dimitrios Koutsonikolas, Josep Jornet, Tommaso Melodia\\
    \textit{Institute for the Wireless Internet of Things, Northeastern University, Boston, MA, U.S.A.}\\
}

\begin{document}

\maketitle

\begin{abstract}
Lunar surface operations impose stringent requirements on wireless communication systems, including autonomy, robustness to disruption, and the ability to adapt to environmental and mission-driven context. While Space-O-RAN provides a distributed orchestration model aligned with 3GPP standards, its decision logic is limited to static policies and lacks semantic integration. We propose a novel extension incorporating a semantic agentic layer enabled by the Model Context Protocol (MCP) and Agent-to-Agent (A2A) communication protocols, allowing context-aware decision making across real-time, near-real-time, and non-real-time control layers. Distributed cognitive agents deployed in rovers, landers, and lunar base stations implement wireless-aware coordination strategies, including delay-adaptive reasoning and bandwidth-aware semantic compression, while interacting with multiple MCP servers to reason over telemetry, locomotion planning, and mission constraints. 
\end{abstract}

\section{Introduction}

Lunar missions~\cite{werkheiser2024lunar} require robust and flexible wireless communication infrastructures capable of functioning in highly dynamic, uncertain, and disconnected environments. Unlike terrestrial deployments, lunar systems must operate with irregular topology, power-constrained mobile nodes, and limited access to Earth. NASA's LunaNet architecture represents a significant advancement in establishing interoperable lunar communications and navigation services~\cite{israel2023lunanet, cooper2020lunanet}, while recent initiatives such as Nokia's lunar cellular network deployments~\cite{israel2023lunanet} and Space-O-RAN frameworks~\cite{baena2025} have explored adaptations of terrestrial technologies to non-terrestrial environments. However, these approaches often result in fragmented and brittle systems that retain assumptions of predefined policies and human-supervised control.

Current lunar communication infrastructures face fundamental limitations in autonomous operation capabilities. The round-trip communication delays of 1.5-2 seconds between Earth and Moon create substantial performance degradation in mission-critical operations, while the dynamic topology challenges in cislunar space highlight the inadequacy of traditional routing protocols designed for stable terrestrial networks~\cite{maaref2023lunar}. Contemporary Space-O-RAN and non-terrestrial network architectures, despite incorporating AI-driven management capabilities through hierarchical Space RIC implementations, remain constrained by static configuration management and rule-based automation insufficient for autonomous decision-making in long-duration lunar missions~\cite{baena2025}.

What is missing is the ability to interpret mission-level intent and context across different layers of the system while maintaining robust inter-agent coordination under wireless constraints. Recent advances in semantic communication and intent-driven networking present compelling solutions to these challenges~\cite{guo2024survey, xin2024semantic}. The emergence of semantic reasoning frameworks capable of integrating spatial positioning data with temporal mission constraints, combined with advances in context-aware network architectures, offers the foundation for networks that can interpret context across multiple system layers~\cite{zeydan2020recent}. The development of Model Context Protocol (MCP) and advances in multi-agent coordination frameworks provide technological building blocks for context-aware infrastructures~\cite{hou2025model}, while Agent-to-Agent communication protocols enable direct semantic coordination between distributed cognitive systems.

Addressing this gap requires integrating a semantic reasoning layer with wireless-aware coordination mechanisms into the network fabric. We propose such a layer using MCP-mediated interactions and A2A communication protocols, enabling agentic behavior directly within the RAN control plane to support autonomous, context-aware lunar network operations that can function effectively under the extreme constraints and challenges unique to lunar environments.

 \begin{figure*}[!h]
    \centering
    \includegraphics[width=0.6\textwidth]{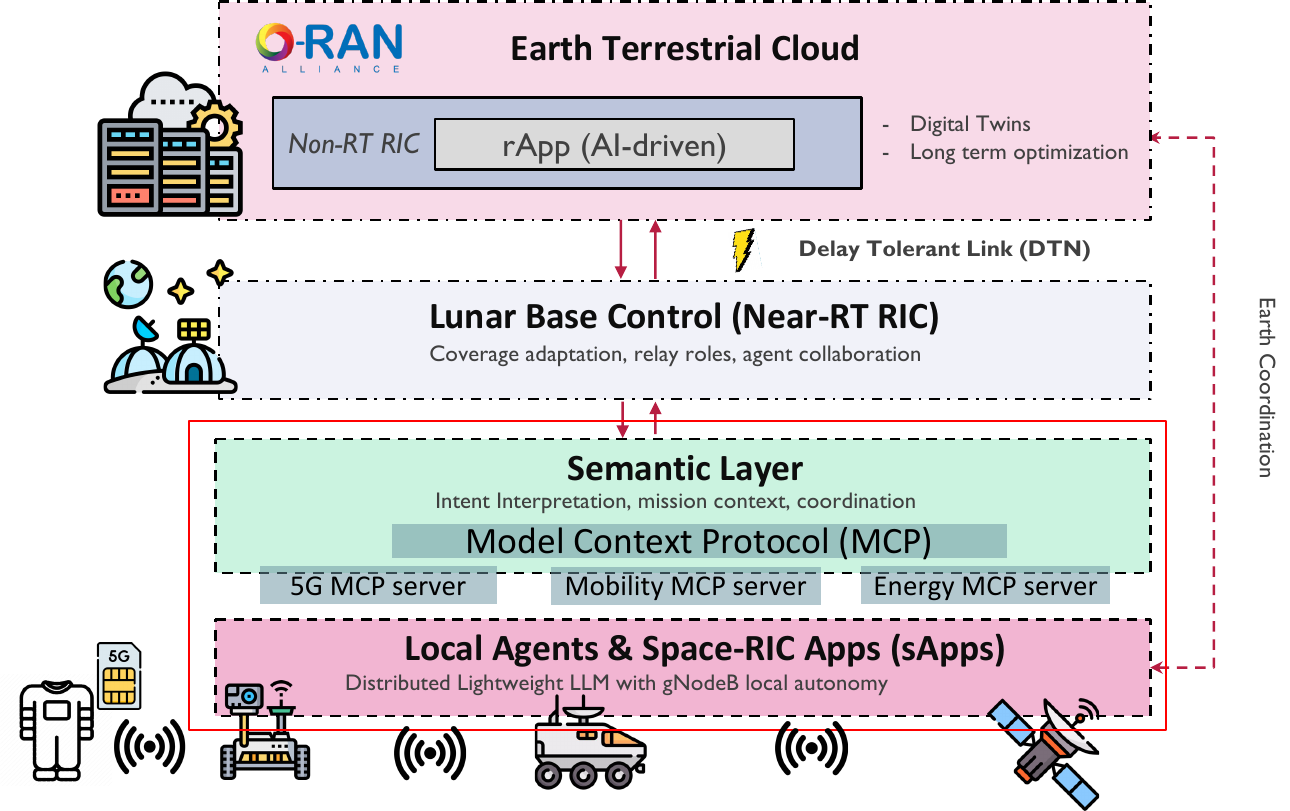}
    \caption{Proposed stack architecture}
    \label{fig:architecture}
\end{figure*}

\section{Architecture}

Our architecture builds upon the Space-O-RAN stack, introducing a semantic control layer over the RIC hierarchy. As shown in Figure~\ref{fig:architecture}, the system spans multiple tiers of control with distributed cognitive agents operating across the entire network fabric.

\subsection{Hierarchical Control Infrastructure}

At the bottom layer, mobile assets such as rovers and landers embed gNodeB functions and lightweight Space-RIC applications (sApps). These include local monitoring modules that observe radio conditions, system load, and agent telemetry. Each mobile asset hosts embedded cognitive agents~\cite{liang2024generative} that interface with local MCP servers for immediate decision-making and context awareness.

The Near-RT RICs, deployed at regional hubs (e.g., relay stations, interim bases), provide edge aggregation and react to local events through policy-driven adaptation, including relay switching and beam steering. These intermediate nodes maintain regional cognitive agents that coordinate between mobile assets and higher-tier controllers.

The lunar base station serves a dual role in this hierarchy: while it typically hosts the Non-RT RIC for the lunar surface domain, it also operates as a cognitive coordination hub. Base station cognitive agents manage network-wide policies, resource allocation, and mission-level coordination. These agents maintain persistent connections to both local Near-RT RICs and Earth-based control systems, serving as intelligent intermediaries that can make autonomous decisions during communication blackouts with Earth.

Above the lunar infrastructure, the Earth-based Non-RT RIC maintains long-term policy, network topology, and digital twin synchronization. Due to the round-trip delays and potential lunar occultation periods, it communicates with lunar components via Delay-Tolerant Networking (DTN) protocols~\cite{fall2003delay} and episodic context updates.

\subsection{Agent-to-Agent Communication Protocols}

The semantic layer introduces autonomous agents distributed across all tiers, employing multiple communication paradigms to ensure robust coordination under varying wireless conditions. Direct agent communication utilizes a lightweight message-passing framework built on top of existing RAN control channels~\cite{li2025glue}, where agents exchange semantic state vectors, policy updates, and coordination messages using structured JSON payloads embedded within O-RAN control messages~\cite{oran2024e2sm}. This approach leverages existing RAN infrastructure while maintaining semantic fidelity across the distributed cognitive network.

Complementing direct communication, each cognitive agent maintains connections to local and remote MCP servers~\cite{hou2025model}, enabling both pull-based queries and push-based notifications. MCP servers expose domain-specific capabilities such as locomotion planning, signal quality estimation, and energy prediction as ontological APIs. Agents can subscribe to context streams from multiple MCP endpoints, enabling reactive coordination based on environmental changes and mission evolution.

The system implements adaptive information dissemination that dynamically adjusts to network conditions. During high-connectivity periods, agents employ push-based updates for real-time coordination with frequent semantic state synchronization across the network. As connectivity degrades, the system transitions to pull-based querying mechanisms that rely on cached context models and predictive pre-fetching strategies. Under severely disrupted connectivity conditions, agents operate autonomously using locally cached semantic models while maintaining periodic bulk synchronization capabilities for eventual state reconciliation.

\subsection{Wireless-Aware Cognitive Operations}

Agent interactions are fundamentally shaped by the underlying wireless characteristics of the lunar environment, requiring sophisticated adaptation mechanisms that account for communication constraints inherent to space-based operations. Cognitive agents adjust their planning horizons based on current and predicted communication delays, implementing delay-adaptive reasoning frameworks where short-term decisions rely on local semantic models while longer-term coordination incorporates delay predictions and confidence intervals.

The system implements intermittency-resilient coordination through probabilistic models of peer availability and opportunistic coordination strategies~\cite{knoblock2020cognitive}. When direct agent communication becomes unavailable, coordination occurs through shared MCP servers acting as semantic message brokers, maintaining system coherence despite communication disruptions. This approach enables continued operation under the challenging connectivity conditions characteristic of lunar surface networks.

Bandwidth constraints necessitate dynamic adjustment of semantic payload granularity based on available communication resources. The system employs bandwidth-aware semantic compression techniques~\cite{yang2024swinjscc} that enable rich context sharing during high-bandwidth periods while implementing semantic summarization and selective information filtering under constrained conditions. Agent decision confidence is modulated by communication link quality metrics, where poor signal conditions reduce inter-agent coordination reliability, causing agents to increase local autonomy and defer non-critical distributed decisions until connectivity improves.

Importantly, cognition is not an application overlay: it is a control layer integrated into the RAN management stack. Through MCP and A2A protocols, an agent can decide whether to reallocate bandwidth, request handovers, or adapt sampling rates not merely due to signal degradation, but based on mission relevance, peer agent recommendations, or human intent parsed from semantic input. This integration enables the network to exhibit intelligent behavior that emerges from the collective decision-making of distributed cognitive agents operating under lunar communication constraints.

\section{Use Case: EVA Incident and Agentic Coordination}

We consider an EVA (extravehicular activity) anomaly scenario demonstrating multi-tier cognitive coordination under lunar communication constraints. An astronaut becomes unresponsive after repeated status pings, detected initially through degraded biometric telemetry.

A nearby rover's cognitive agent detects the anomaly through semantic behavior model analysis and immediately initiates Agent-to-Agent communication with other mobile assets within direct radio range, broadcasting a semantic alert containing the anomaly classification, location uncertainty, and required assistance level. Due to lunar terrain shadowing, the rover implements adaptive coordination strategies, caching critical context locally while attempting to establish relay paths through intermediate assets.

A secondary rover positioned on higher terrain receives the A2A alert and acts as a communication relay, with its cognitive agent evaluating relay capacity using MCP-provided link quality predictions. The relay agent coordinates resource allocation, dedicating higher bandwidth to emergency traffic while degrading non-critical data flows. Meanwhile, the lunar base station cognitive agent, upon receiving relayed information through DTN buffering, immediately reconfigures network priorities and instructs Near-RT RIC components to reallocate spectrum resources toward the incident area.

Given the RTT delay to Earth during this scenario, the base station agent makes autonomous decisions while preparing detailed situation reports for Earth-based Non-RT RIC systems. The rescue rover employs MCP-driven locomotion planning, continuously adapting its trajectory based on real-time wireless quality feedback from base station agents, switching to intermittency-resilient mode when wireless conditions degrade due to terrain occlusion.

Throughout the incident, agents dynamically select communication protocols based on wireless conditions, transitioning from rich A2A semantic exchange with frequent MCP context updates under high connectivity to compressed semantic payloads with selective MCP querying under moderate connectivity, and finally to cached decision-making with opportunistic bulk synchronization under poor connectivity conditions. This demonstrates how cognitive agents coordinate across multiple domains, reasoning, mobility, and radio resource orchestration,while adapting their communication strategies to lunar wireless constraints, ensuring mission continuity even under partial observability and intermittent connectivity.

\section{Conclusions and Future Work}

This work proposes an agentic control architecture that embeds semantic cognition into the orchestration fabric of wireless lunar networks. The integration of lightweight reasoning agents within the O-RAN control stack, enhanced with Agent-to-Agent communication protocols and MCP-mediated interactions, enables decentralized, policy-compliant adaptation to evolving mission conditions without requiring continuous backhaul availability.

Several critical challenges emerge from this wireless-aware cognitive architecture that require systematic investigation. The challenge of resource-constrained cognition presents a fundamental trade-off between cognitive capability and power consumption, particularly acute for mobile assets operating under limited energy budgets. Running real-time inference and memory coordination on radiation-hardened, resource-constrained edge platforms demands comprehensive optimization of semantic models and computation pipelines.

The scalability of Agent-to-Agent protocol implementation introduces significant concerns as cognitive agent populations increase, where A2A communication overhead may saturate available control channels, necessitating the development of hierarchical agent clustering methodologies and selective communication strategies to maintain network scalability while preserving coordination effectiveness. Wireless-induced cognitive drift represents a novel challenge where poor communication conditions can lead to semantic model divergence across distributed agents, requiring innovative approaches to distributed consensus and conflict resolution in semantic reasoning systems.

Safety verification requirements become increasingly complex when agent policies must operate under communication uncertainty, encompassing developing runtime monitors capable of detecting semantic drifts and wireless-induced coordination failures while ensuring graceful degradation when agent coordination capabilities are compromised. The integration of Delay-Tolerant Networking with semantic reasoning systems presents unique consistency challenges, where asynchronous state propagation over DTN links must preserve consistency guarantees across distributed agents operating under non-uniform time horizons.

Current research efforts focus on deploying containerized semantic agents over emulated lunar mesh topologies, utilizing synthetic telemetry data to systematically validate Agent-to-Agent coordination protocols, memory synchronization mechanisms, and trajectory adaptation algorithms under realistic wireless propagation models. Future work will explore federated cognition schemes that enable agents to collaboratively learn environment dynamics and optimize policies without raw data exchange, wireless-semantic co-design for joint optimization of resource allocation and reasoning workloads, and cross-layer security frameworks addressing vulnerabilities introduced by A2A communication and MCP-mediated interactions. 

\bibliographystyle{IEEEtran}
\bibliography{main}


\end{document}